\documentclass[a4paper,twoside,reqno]{bjp}

\usepackage{graphicx}
\usepackage{cite}
\usepackage{amssymb,amsmath,amscd,amsthm}
\usepackage{times}

\usepackage{geometry}\allowdisplaybreaks

\theoremstyle{plain}\newtheorem{thm}{Theorem}
\theoremstyle{remark}\newtheorem*{rem}{Remark}

\newcommand\gGL{\text{\normalfont GL}}
\newcommand\gO{\text{\normalfont O}}
\newcommand\gSp{\text{\normalfont Sp}}
\newcommand\gSO{\text{\normalfont SO}}
\newcommand\gPin{\text{\normalfont Pin}}
\newcommand\gSpin{\text{\normalfont Spin}}
\newcommand\gCl{\text{\normalfont Cl}}

\newcommand\lo{\mathfrak{o}}

\begin{document}

\title{Proof of the orthogonal--Pin duality}

\runningheads{Proof of the orthogonal--Pin duality}{K. Neerg\aa rd}

\begin{start}{

\author{K. Neerg\aa rd}{}

\index{Neerg\aa rd, K.}

\address{Fjordtoften 17, 4700 N\ae stved, Denmark}{}

\received{}

}

\begin{Abstract}
  This article contains the proof of a theorem on orthogonal--Pin
  duality that was cited without proof in a previous article in this
  journal.
\end{Abstract}

\begin{KEY}
  Orthogonal Lie algebra, orthogonal group, Spin and Pin groups,
  duality.
\end{KEY}

\end{start}

\section{\label{sec:intr}Introduction}

In a previous article in this journal~\cite{ref:Nee21}, I cited a
theorem of dual Fock space representations of an orthogonal Lie
algebra and a Pin group, but the space limits of that article as a
conference contribution did not allow to include the proof. The
present paper serves to render the proof, which is found so far only
in a preprint~\cite{ref:Nee20a}, accessible in a journal. For the
context of the theorem, which is Theorem~\ref{th:oP} below,
see~\cite{ref:Nee21,ref:Nee20b}. The proof is given in
Section~\ref{sec:oP}. Sections~\ref{sec:term}--\ref{sec:pin} provide
necessary preliminaries. I thus specify my terminology in
Section~\ref{sec:term} and define the fermion Fock space in
Section~\ref{sec:fock}. In Section~\ref{sec:lie}, I review the
constructions on this space of a number conserving and a commuting
number non-conserving representation of orthogonal Lie
algebras~\cite{ref:Nee19}, and in Section~\ref{sec:pin}, I extend the
number non-conserving representation to a representation of a Pin
group as defined by Atiyah, Bott and Shapiro~\cite{ref:Ati64}.
Section~\ref{sec:dis} extends the discussion in~\cite{ref:Nee21} by
relating my result to contemporary work on Fock space dualities, and
Section~\ref{sec:sum} provides a summary.

\section{\label{sec:term}Terminology and notation}

While the theorem to be proved is one of linear algebra, I use the
terminology and notation of quantum mechanics. Throughout, the base
field is that of the complex numbers. Members of a finite-dimensional
vector space $V$ are written as Dirac kets
$|i \rangle$~\cite{ref:Dir30} and sometimes called states. Members of
the dual space $V^*$ of linear forms on $V$ are written as Dirac bras
$\langle i|$, and $\langle i| (|j \rangle)$ as
$\langle i | j \rangle$. When $(|i \rangle, i = 1, \dots, n)$ is a
basis for a vector space $V$ of finite dimension $n$, its dual basis
$(\langle i|, i = 1, \dots, n)$ for $V^*$ is defined by
$\langle i | j \rangle = \delta_{ij}$ and vice versa, and when a ket
and a bra have identical labels, they are understood to be
corresponding members of dual bases. In particular, the identity on
$V$ may thus be written $\sum_i |i \rangle \langle i|$, where
$|i \rangle$ scans an arbitrary basis for $V$. Bilinear forms on $V$
are written as bras $\langle b |$, and
$\langle b| (|i \rangle,|j \rangle)$ as $\langle b| i j \rangle$.
Bilinear forms on $V^*$ are written as kets $|b \rangle$, and
$|b \rangle (\langle i |,\langle j |)$ as $\langle i j | b \rangle$.
When $\langle b |$ is non-degenerate, a dual bilinear form
$| b \rangle$ is defined by
$\sum_k \langle b| j k \rangle \langle i k |b \rangle = \langle i | j
\rangle$, which renders $|b \rangle$ non-degenerate, and vice versa.
The bilinear form $| b \rangle$ is symmetric if and only if
$\langle b|$ is symmetric.

\section{\label{sec:fock}Fock space}

The theorem to be proved concerns group representations on the
\textit{Fock space} of multiple kinds of fermions sharing a
single-kind state space. I therefore introduce this concept. Consider
$k$ \textit{kinds} of fermions inhabiting a common \textit{single-kind
  state space} $\mathcal S$ of dimension $d$. Examples are spin up and
down electrons in an atomic shell with orbital angular momentum $l$
($d = 2l + 1$, $k = 2$), spin up and down nucleons in a nuclear shell
with orbital angular momentum $l$ ($d = 2l + 1$, $k = 4$), nucleons in
a nuclear shell with orbital and spin angular momenta coupled to total
angular momentum $j$ ($d = 2j + 1$, $k = 2$). In terms of a state
space $\mathcal K$ with a basic state $|\tau \rangle$ for each fermion
kind $\tau$, one can define, corresponding to every state
$|p \rangle \otimes |\tau \rangle$ in $\mathcal S \otimes \mathcal K$,
a \textit{creation operator} $a_{p\tau}^\dagger$, and corresponding to
every member $\langle p| \otimes \langle \tau|$ of
$\mathcal S^* \otimes \mathcal K^*$, an \textit{annihilation operator}
$a_{p\tau}$. By definition, these operators obey
\begin{equation}\label{eq:anti}
  \{ a_{p\tau} , a_{q\upsilon}^\dagger \}
  = \langle p | q \rangle \langle \tau | \upsilon \rangle , \quad
  \{ a_{p\tau} , a_{q\upsilon} \}
  = \{ a_{p\tau}^\dagger , a_{q\upsilon}^\dagger \} = 0
\end{equation}
in terms of the anticommutator $\{\cdot,\cdot\}$. Formally,
$\mathcal S \otimes \mathcal K$ is just a $dk$-dimensional vector
space, and the relations~\eqref{eq:anti} are those of exterior and
interior multiplication by members of $\mathcal S \otimes \mathcal K$
and $\mathcal S^* \otimes \mathcal K^*$ on the \textit{exterior
  algebra}~\cite{ref:Jac80} on $\mathcal S \otimes \mathcal K$ .

I call linear combinations of the creation and annihilation operators
\textit{field operators} and denote by $\mathcal F$ the space of such
operators. They generate the \textit{Clifford algebra}
$\gCl(2dk)$~\cite{ref:Cli78} of the anticommutator product in
$\mathcal F$. The members of $\gCl(2dk)$ act on a space $\Phi$ that is
isomorphic as a vector space to the subalgebra of $\gCl(2dk)$
generated by the creation operators, the \textit{spinor space} of
$\gCl(2dk)$~\cite{ref:Bra35}. \emph{This} is the Fock space. One state
$|\rangle$ in $\Phi$ corresponds to the unit member of the subalgebra,
and the action of $\gCl(2dk)$ on $\Phi$ is given by the convention
that every annihilation operator kills $|\rangle$, which may thus be
seen as a \textit{vacuum state}. This renders $\gCl(2dk)$ identical to
the algebra of linear transformations of $\Phi$. One may alternatively
identify $\Phi$ with the exterior algebra on
$\mathcal S \otimes \mathcal K$, which renders the creation and
annihilation operators identical to the operators of exterior and
interior multiplication by members of $\mathcal S \otimes \mathcal K$
and $\mathcal S^* \otimes \mathcal K^*$. Note that I \emph{do not}
introduce any Hermitian inner product on $\Phi$, so despite the
notation, $a_{p\tau}$ and $a_{p\tau}^\dagger$ should not be seen as a
Hermitian conjugates.

\section{\label{sec:lie}Fock space representations of orthogonal Lie
  algebras}

I refer to~\cite{ref:Jac62} for the basic theory of orthogonal Lie
algebras. The \textit{orthogonal Lie algebra} $\lo(n)$ is the Lie
algebra of infinitesimal linear transformations $x$ of a vector space
$V$ of finite dimension $n$ that preserves a non-degenerate, symmetric
bilinear form $\langle b |$ on $V$ in the sense that
\begin{equation}
  \langle b | (x |i \rangle \otimes |j \rangle
  + |i \rangle \otimes x |j \rangle) = 0
\end{equation}
for every $|i \rangle , |j \rangle \in V$. Different $\langle b |$
make isomorphic Lie algebras. A convenient choice is 
\begin{equation}\label{eq:b}
  \langle b | i j \rangle = \delta_{i+j,0}
\end{equation}
relative to a basis $(|i \rangle, i = -\Omega , \dots , \Omega)$ with
$|0 \rangle$ omitted when $n$ is even, where
$\Omega = \lfloor n/2 \rfloor$. In terms of basic elements
\begin{equation}
  e_{ij} = |i \rangle \langle j|
\end{equation}
of the space of linear transformations of $V$, where $|i \rangle$ and
$\langle j|$ belong to dual bases for $V$ and $V^*$, the Lie algebra
$\lo(n)$ defined by~\eqref{eq:b} is spanned by the transformations
\begin{equation}\label{eq:onbas}
  \bar e_{ij} = e_{ij} - e_{-j,-i}
\end{equation}
with $i + j > 0$. Its finite-dimensional irreducible representations
are characterised by \textit{highest weights}
$\Lambda = (\Lambda_1 , \dots , \Lambda_\Omega)$, where each
$\Lambda_i$ is the module eigenvalue of $\bar e_{ii}$ on a
\textit{highest weight vector} killed by the module action of every
$\bar e_{ij}$ with $i > j$. (Isomorphic representations are considered
identical unless otherwise specified.) The highest weight vector is
unique within normalisation and the entire irreducible module is
generated from it by polynomials in the module actions of the basic
elements~\eqref{eq:onbas} of the Lie algebra. The highest weight
components $\Lambda_i$ are either integral or half-integral and obey
\begin{equation}
  \begin{array}{ll}
    0 \le \Lambda_1 \le \Lambda_2 \le \dots \le \Lambda_\Omega & 
      \text{when $n$ is odd ,} \\
    0 \le |\Lambda_1| \le \Lambda_2 \le \dots \le \Lambda_\Omega &
      \text{when $n$ is even .}
  \end{array}
\end{equation}

Highest weights are conveniently visualized by generalized Young
diagrams~\cite{ref:Nee20b}. For example, the diagram
\begin{equation}
  \includegraphics{young16}
\end{equation}
describes the highest weight $\Lambda = (-3/2,5/2,9/2)$. Rows of zero
length are omitted. In particular the diagram is empty if every
$\Lambda_i$ is zero. In the case of $\lo(2)$, the edge whence the
single row extends must be specified. By convention, $-\Lambda$ shall
denote for even $n$ the highest weight obtained from $\Lambda$ by
changing the sign of $\Lambda_1$.

\begin{rem}
  The Lie algebras $\lo(1)$, $\lo(2)$ and $\lo(4)$ are special.
  $\lo(1)$ is 0-dimensional, and its only irreducible representation
  consist of the operator 0 on a 1-dimensional vector space. It
  concurs with the systematics to assign to this representation
  $\Lambda = ()$ (the 0-tuple) and an empty diagram. $\lo(2)$ is
  1-dimensional. Its irreducible representations are described by an
  arbitrary complex number $\Lambda_1$, but only those with an
  integral or half-integral $\Lambda_1$ occur in the sequel. $\lo(4)$
  is isomorphic to $\lo(3) \oplus \lo(3)$, and its irreducible modules
  with highest weights $(\Lambda_1,\Lambda_2)$ are products of
  irreducible $\lo(3)$ modules with single highest weight components
  $(\Lambda_2 \pm \Lambda_1)/2$~\cite{ref:Nee19}.
\end{rem}

Every $\lo(n)$ is the Lie algebra of the \textit{orthogonal group}
$\gO(n)$ of linear transformation $g$ of $V$ that preserve $\langle
b|$ in the sense that
\begin{equation}
  \langle b | (g |i \rangle \otimes g |j \rangle)
  = \langle b | (|i \rangle \otimes |j \rangle)
\end{equation}
for every $|i \rangle , |j \rangle \in V$. This implies
$\det g = \pm 1$. The elements $g \in \gO(n)$ with $\det g = 1$ form
the subgroup $\gSO(n)$ of \textit{proper} orthogonal transformations.

The Fock space $\Phi$ carries two commuting faithful representations
of orthogonal Lie algebras~\cite{ref:Nee19}. The first one is a
representation of $\lo(d)$ defined by
\begin{equation}\label{eq:d}
  x \mapsto \sum_{pq\tau}
    a_{p\tau}^\dagger \langle p | x | q \rangle a_{q\tau}
\end{equation}
for every $x \in \lo(d)$. The second one is a representation of
$\lo(2k)$ given by
\begin{multline}\label{eq:2k}
    \bar e_{\tau\upsilon} \mapsto
      \sum_p a_{p\tau} a_{p\upsilon}^\dagger 
      - \delta_{\tau\upsilon} \frac d 2, \quad
    \bar e_{\tau,-\upsilon} \mapsto \sum_{pq}
      \langle b | p q \rangle
      a_{p\tau} a_{q\upsilon} ,  \\
    \bar e_{-\tau,\upsilon} \mapsto \sum_{pq}
      a^\dagger_{p\tau} a^\dagger_{q\upsilon}
      \langle p q | b \rangle .
\end{multline}
in terms of the bilinear form $\langle b|$ that defines $\lo(d)$. For
$\lo(2k)$, the bilinear form~\eqref{eq:b} is implicit. In~\eqref{eq:d}
and \eqref{eq:2k}, summation indices $p$ and $q$ scan the labels of a
basis for $\mathcal S$ and the indices $\tau$ and $\upsilon$ scan the
set of fermion kinds. In~\cite{ref:Nee19}, these representations are
called number conserving and number non-conserving, respectively. By a
calculation of characters, I prove in~\cite{ref:Nee19} the following.

\begin{thm}[$\lo(d)$--$\lo(2k)$ duality]\label{th:oo}
  The fermion Fock space $\Phi$ has the decomposition
\begin{equation}\label{eq:oo}
  \Phi = \bigoplus {\textup X}_\lambda \otimes \Psi_\mu ,
\end{equation}
where $\textup{X}_\lambda$ and $\Psi_\mu$ carry representations of
$\lo(d)$ and $\lo(2 k)$. The summation in~\eqref{eq:oo} runs over all
pairs $(\lambda,\mu)$ of highest weights such that the $\lambda$ Young
diagram and a reflected and rotated copy of the $\mu$ Young diagram
fill a $d/2 \times k$ frame without overlap as in the following
example, where $(d,k) = (11,4)$, $\lambda = (1,2,2,3,4)$ and
$\mu = (1/2,3/2,7/2,9/2)$.

{\centering\includegraphics{young30}\par}

If the border between the diagrams hits the bottom of the frame (which
is possible only when $d$ is even),
${\textup X}_\lambda = \bar{\textup X}_\lambda \oplus \bar{\textup
  X}_{-\lambda}$, where $\bar{\textup X}_{\pm\lambda}$ are irreducible
with highest weights $\pm\lambda$, and $\Psi_\mu$ is irreducible with
highest weight $\mu$. If the border hits the left edge (as in the
example), ${\textup X}_\lambda$ is irreducible with highest weight
$\lambda$, and $\Psi_\mu = \bar\Psi_\mu \oplus \bar\Psi_{-\mu}$, where
$\bar\Psi_{\pm\mu}$ are irreducible with highest weights $\pm\mu$.
\end{thm}

The direct product of an $\lo(d)$ highest weight vector of
$\bar{\textup X}_{\pm\lambda}$ or ${\textup X}_\lambda$ and an
$\lo(2k)$ highest weight vector of $\Psi_\mu$ or $\bar\Psi_{\pm\mu}$
is an $\lo(d) \oplus \lo(2k)$ highest weight state of the product
module. In~\cite{ref:Nee20b}, I construct these states. To this end, I
define modified $\lo(d)$ Young diagrams of maximal depth $d$ whose
rows are labelled
$p = \lfloor d/2 \rfloor , \lfloor d/2 \rfloor - 1 ,\dots$ from the
top with $p = 0$ omitted when $d$ is even. The columns are labeled
$\tau = 1,2,\dots$ from the left. For each such diagram $D$,\, I set
\begin{equation}\label{eq:phiD}
  \phi_D = \left( \prod_{p\tau \in D}
    a^\dagger_{p\tau} \right)  |\rangle ,
\end{equation}
where the product runs over the cells of $D$ with each cell labelled
by its row $p$ and column $\tau$. The order of the factors
$a^\dagger_{p\tau}$ is immaterial. When $d$ is even and
$\lambda_1 > 0$, I consider besides the $\lambda$ diagram the diagram
obtained by moving its bottom row one step down. For example, for
$d=6$ and $\lambda = (2,4,5)$, the change is

\begin{equation}\label{eq:exceptD}\text{
\raisebox{-20bp}{\includegraphics{young4}}
$\longrightarrow$
\raisebox{-30bp}{\includegraphics{young33}}.
}\end{equation}

I then set $D = \lambda$ for the original diagram and $D = -\lambda$
for the changed diagram. Then $\phi_{\pm\lambda}$ are
$\lo(d) \oplus \lo(2k)$ highest weight states of
$\bar{\textup X}_{\pm\lambda} \otimes \Psi_\mu$. Otherwise if
$\tilde\lambda_1$ is the depth of the first column, I extend this
column to the depth $d - \tilde\lambda_1$. For example, for $d=5$ and
$\lambda = (2,4)$, the change is

\begin{equation}\text{
\raisebox{-10bp}{\includegraphics{young34}}
$\longrightarrow$
\raisebox{-20bp}{\includegraphics{young35}}.
}\end{equation}

I set $D = \lambda$ for the original diagram and $D = \lambda'$ for
the changed diagram. Then $\phi_\lambda$ and $\phi_{\lambda'}$ are
$\lo(d) \oplus \lo(2k)$ highest weight states of
${\textup X}_\lambda \otimes \bar\Psi_{\pm\mu}$.

\section{\label{sec:pin}Fock space representations of $\gPin(2k)$}
 
Definitions of a groups $\gPin(n)$ vary in the literature, but every
definition gives within isomorphism the same subgroups $\gSpin(n)$
(see below), isomorphic for $n \ge 3$ to the universal covering
groups~\cite{ref:Wey39} of $\gSO(n)$. I adopt the definition due to
Atiyah, Bott and Shapiro~\cite{ref:Ati64} as naturally generalized
from the real to the complex case by Goodman and
Wallach~\cite{ref:Goo98}. Since I am concerned only with the groups
$\gPin(2k)$, I shall not consider the case when $n$ is odd.

Atiyah, Bott and Shapiro construct $\gPin(2k)$ within the Clifford
algebra $\gCl(2k)$, which is obtained by setting $d = 1$ in
section~\ref{sec:fock}. I denote in this case the space $\mathcal F$
of field operators by $\mathcal F_1$ and omit the index 1 in
$a_{1\tau}$ and $a_{1\tau}^\dagger$. Then $\gPin(2k)$ is the group of
all products of elements $\alpha \in \mathcal F_1$ such that
$\alpha^2 = -1$. An operator $\iota$ on $\gCl(2k)$ is given by
$\alpha \mapsto -\alpha$ for $\alpha \in \mathcal F_1$, and an
operator $\theta$ on $\gCl(2k)$ inverts the order of the factors in
any product of members of $\mathcal F_1$. The bilinear form
$\{\alpha,\beta\}$ defines a realization of $\gO(2k)$ on
$\mathcal F_1$, which I identify with $\gO(2k)$. It can be shown that
for every $g \in \gPin(2k)$ such that
\begin{equation}
  g (\iota \theta g) = 1 ,
\end{equation}
the transformation
\begin{equation}\label{eq:u}
  u: \quad \alpha \mapsto (\iota g) \alpha (\iota \theta g) , \quad
   \forall \alpha \in \mathcal F_1 ,
\end{equation}
belongs to this $\gO(2k)$ and that the map
$\gPin(2k) \rightarrow \gO(2k): g \mapsto u$ is
surjective~\cite{ref:Ati64,ref:Goo98}. Evidently, elements
$\pm g \in \gPin(2k)$ map to the same $u$, so $g \mapsto u$ provides a
\emph{double covering} of $\gO(2k)$ by $\gPin(2k)$. The Lie algebra of
$\gPin(2k)$ is then isomorphic to $\lo(2k)$, and I identify it with
$\lo(2k)$

The products of an even number of factors $\alpha \in \mathcal F_1$
with $\alpha^2 = -1$ form a subgroup $\gSpin(2k)$ of $\gPin(2k)$ which
double covers $\gSO(2k)$ and is the maximal connected subset of
$\gPin(2k)$. The Lie algebra of $\gSpin(2k)$ is that of $\gPin(2k)$,
that is, $\lo(2k)$. The relation~\eqref{eq:u} leads to a defining
relation
\begin{equation}\label{eq:x}
  (\iota x) \alpha + \alpha (\iota \theta x) = 0, \quad
    \forall \alpha \in \mathcal F_1 ,
\end{equation}
for the members $x$ of $\lo(2k)$. It is easily verified that
\eqref{eq:x} holds when $x$ is any commutator $[\beta,\gamma]$ of
field operators $\beta,\gamma \in \mathcal F_1$. The span of the set
of these commutators is exactly the span of the set of operators on
the right in~\eqref{eq:2k} for $d = 1$, so it exhausts $\lo(2k)$.
Since the representation~\eqref{eq:2k} is faithful, one therefore gets
a representation $\rho$ of $\lo(2k)$ on the Fock space $\Phi$ by
composing the map~\eqref{eq:2k} with the inverse map for $d = 1$.

I now choose basic vectors $|f \rangle, |g \rangle, \dots$ in
$\mathcal S$ such that $\langle b | f g \rangle = \delta_{fg}$ and set
\linebreak
$\alpha_f = \sum_\tau (u_\tau a_{f\tau} + v_\tau a_{f\tau}^\dagger)$
when $\alpha = \sum_\tau (u_\tau a_\tau + v_\tau a_\tau^\dagger)$ with
numeric coefficients $u_\tau$ and $v_\tau$. The map $\rho$ is then
given by
\begin{equation}
  \rho: \quad
  [\alpha,\beta] \mapsto \sum_f [\alpha_f,\beta_f] , \quad
  \forall \alpha,\beta \in \mathcal F_1 .
\end{equation}
This evidently expands to a representation of $\gSpin(2k)$ given by
\begin{equation}\label{eq:rhoSpin}
  \rho: \quad g \mapsto \prod_f g_f ,
\end{equation}
where $g_f$ is obtained by substituting every
$\alpha \in \mathcal F_1$ by $\alpha_f$ in the expression for $g$ as a
member of $\gCl(2k)$. Indeed, since $g$ is the product of an even
number of factors $\alpha$, the factors in~\eqref{eq:rhoSpin} commute.
To get a representation of $\gPin(2k)$, it then suffices to construct
$\rho(\alpha)$ for one $\alpha \in \mathcal F_1$ with $\alpha^2 = -1$
such that (i) for every $g,g' \in \gSpin(2k)$ such that
$\alpha g = g' \alpha$ the identity
$\rho(\alpha) \rho(g) = \rho(g') \rho(\alpha)$ holds, and (ii)
$\rho(\alpha)^2 = \rho(-1)$. Because $\gSpin(2k)$ is connected, the
condition (i) is equivalent to (i')
$\alpha x = x' \alpha \Rightarrow \rho(\alpha) \rho(x) = \rho(x')
\rho(\alpha)$ for every $x,x' \in \lo(2k)$.

This condition holds if there is a map $\beta \mapsto \beta'$ of
$\mathcal F_1$ into itself such that $\alpha \beta = \beta' \alpha$
for $\beta \in \mathcal F_1$ and
$\rho(\alpha) \beta_f = s \beta'_f \rho(\alpha)$ for every $\beta$ and
$f$ with a fixed sign $s = +$ or $-$. The map $\beta \mapsto \beta'$
exists as
$\beta \mapsto - \alpha \beta \alpha = - \beta - \{\alpha,\beta\}
\alpha$. Further, $\alpha_f \beta_g = - \beta_g \alpha_f$ for
$f \ne g$ and $\alpha_f \beta_f = s \beta'_f \alpha_f$, so
$\rho(\alpha) \beta_f = s \beta'_f \rho(\alpha)$ holds with
\begin{equation}\label{eq:rhoal}
  \rho(\alpha) = c \prod_f \alpha_f
\end{equation}
and $s = (-)^{d-1}$, where $c$ is a numeric factor. Because the
factors $\alpha_f$ in \eqref{eq:rhoal} anticommute, their order is
immaterial. To be specific, I set
$f = \Omega, \Omega - 1, \dots , -\Omega$ with 0 omitted when $d$ is
even and assume this order of the factors in \eqref{eq:rhoal}.

As to the condition (ii), first notice
\begin{multline}\label{eq:rho-1}
  \rho(-1) = \rho(\exp i \pi [a_1,a_1^\dagger])
   = \exp \rho(i \pi [a_1,a_1^\dagger]) \\
    = \exp i \sum_f [a_{f1},a_{f1}^\dagger]
    = \prod_f \exp i \pi [a_{f1},a_{f1}^\dagger] = (-1)^d .
\end{multline}
Further, by $\alpha_f \alpha_g = - \alpha_g \alpha_f$ for $f \ne g$
and $\alpha_f^2 = -1$, the expression \eqref{eq:rhoal} gives
$\rho(\alpha)^2 = (-)^{d+\Omega} c^2$, so (ii) holds when $c$ is a
square root of $(-1)^\Omega$. I choose $c = i^\Omega$.

I now set
\begin{equation}
  \alpha = a_1^\dagger - a_1 ,
\end{equation}
whence follows
\begin{equation}
  a'_1 = - a^\dagger_1 , \quad (a^\dagger)'_1 = - a_1 , \quad
  a'_\tau = - a_\tau ,
  \quad (a^\dagger_\tau)' = - a^\dagger_\tau , \quad \tau > 1 ,
\end{equation}
\newpage and with $\sigma =\rho(\alpha)$,
\begin{multline}\label{eq:sigf}
  \sigma a_{f1} = (-)^d a^\dagger_{f1} \sigma , \quad
  \sigma a^\dagger_{f1} = (-)^d a_{f1} \sigma , \\
  \sigma a_{f\tau} = (-)^d a_{f\tau} \sigma , \quad
  \sigma a^\dagger_{f\tau} = (-)^d a^\dagger_{f\tau} \sigma , \quad
    \tau > 1, \forall f .
\end{multline}
Further,
\begin{equation}\label{eq:sigvacf}
  \sigma |\rangle
  = i^\Omega \left( \prod_f a^\dagger_{f1} \right) |\rangle
\end{equation}
with the ordering of the indices $f$ as in~\eqref{eq:rhoal}. The
operator $\sigma$ is similar but not identical to the operator denoted
by this symbol in~\cite{ref:Nee20b}. It shares, in particular, with
the latter the property that it commutes with the module action of
every $g \in \gSO(d)$ by the representation of $\gO(d)$ on $\Phi$
defined in~\cite{ref:Nee20b} and anticommutes with the module action
of every $g \in \gO(d) \setminus \gSO(d)$. The restriction to $\lo(d)$
of this representation is given by~\eqref{eq:d}. The commutation or
anticommutation property of the present $\sigma$ is an easy
consequence of the fact that the product in \eqref{eq:rhoal} acquires
a factor $\det g$ by the change of basis
$|f \rangle \mapsto g |f \rangle$.

Next I change the basis for $\mathcal S$ to the set of vectors $|p
\rangle, |q \rangle, \dots$ given by
\begin{equation}\label{eq:t}
  |\pm p \rangle = \sqrt{\tfrac12} (|f \rangle \pm i |-f \rangle)
    \quad \text{for $p = f > 0$} , \quad
  |p \rangle = |f \rangle \quad \text{for $p = f = 0$} .    
\end{equation}
Then $\langle b |p q \rangle = \delta_{p+q,0}$, so the description of
highest weight vectors in section~\ref{sec:lie} applies. The dual
basis is given by
\begin{equation}
  \langle \pm p| = \sqrt{\tfrac12} (\langle f| \mp i \langle -f|)
    \quad \text{for $p = f > 0$} , \quad
  \langle p| = \langle f| \quad \text{for $p = f = 0$} ,  
\end{equation}
so \eqref{eq:sigf} becomes
\begin{multline}\label{eq:sigp}
  \sigma a_{p1} = (-)^d a^\dagger_{-p,1} \sigma , \quad
  \sigma a^\dagger_{p1} = (-)^d a_{-p,1} \sigma , \\
  \sigma a_{p\tau} = (-)^d a_{p\tau} \sigma , \quad
  \sigma a^\dagger_{p\tau} = (-)^d a^\dagger_{p\tau} \sigma , \quad
    \tau > 1, \forall f .
\end{multline}
Since the inverse of the transformation~\eqref{eq:t} has determinant
$i^\Omega$, equation~\eqref{eq:sigvacf} becomes
\begin{equation}\label{eq:sigvacp}
  \sigma |\rangle
  = (-)^\Omega \left( \prod_p a^\dagger_{p1} \right) |\rangle
\end{equation}
with the indices $p$ ordered as the indices $f$ in~\eqref{eq:rhoal}.

\section{\label{sec:oP}$\lo(d)$-$\gPin(2k)$ duality}

Now consider the action of $\sigma$ on the highest weight
states~\eqref{eq:phiD}. With an appropriate ordering of the factors
$a_{p\tau}^\dagger$ in~\eqref{eq:phiD}, one can write
\begin{equation}
  \phi_D = \left( \prod_{p \in R} a^\dagger_{p1} \right) \phi_{D'} ,
\end{equation}
where $R$ is the set of labels $p$ of the rows in $D$, taken in the
product to be ordered from the top, and $D'$ is $D$ without the first
column. By~\eqref{eq:sigp} and~\eqref{eq:sigvacp}, one gets
\begin{equation}
  \sigma \phi_{D'}
    = (-)^\Omega \left( \prod_p a^\dagger_{p1} \right) \phi_{D'} ,
\end{equation}
whence by~\eqref{eq:sigp},
\begin{equation}
  \sigma \phi_D
  = (-)^{|R|d+\Omega} \left( \prod_{p \in R} a_{-p,1} \right)
    \left(\prod_p a^\dagger_{p1} \right) \phi_{D'} .
\end{equation}
For every $D$ in section~\ref{sec:lie} except the one on the right
in~\eqref{eq:exceptD}, this gives
\begin{equation}\label{eq:sigD}
  \sigma \phi_D = (-)^{|R|+\Omega} 
     \left( \prod_{p \in \complement R} a_{-p,1}^\dagger \right) \phi_{D'} ,
\end{equation}
where the complement $\complement R$ is relative to the set of
possible indices $p$ and the product is taken in the order of
decreasing $-p$.

When $d$ is even and $\lambda_1 > 0$, equation~\eqref{eq:sigD} becomes
\begin{equation}\label{eq:sigphilam}
  \sigma \phi_\lambda = \phi_\lambda .
\end{equation} 
Let $\phi_\lambda = |\chi \rangle \otimes |\psi \rangle$ with
$|\chi \rangle \in \textup{X}_\lambda$ and
$|\psi \rangle \in \Psi_\mu$. Since $\sigma$ commutes with the
representation of $\lo(d)$ on $\Phi$ and it was shown that for every
$x \in \lo(2k)$ there is an $x' \in \lo(2k)$ such that
$\sigma \rho(x) = \rho(x') \sigma$, it follows
from~\eqref{eq:sigphilam} that $|\chi \rangle \otimes \Psi_\mu$ is
invariant to $\sigma$. The space $|\chi \rangle \otimes \Psi_\mu$ then
carries a representation of $\gPin(2k)$, which is irreducible because
$|\chi \rangle \otimes \Psi_\mu$ is irreducible as an $\lo(2k)$
module. By the natural identification of
$|\chi \rangle \otimes \Psi_\mu$ with $\Psi_\mu$, this representation
becomes an irreducible representation of $\gPin(2k)$ on $\Psi_\mu$ and
$|\psi \rangle$ a module eigenvector of $\alpha$ with eigenvalue 1.

A similar calculation gives
\begin{equation}
  \sigma \phi_{-\lambda} = - \phi_{-\lambda} .
\end{equation}
It follows again that $\Psi_\mu$ carries an irreducible representation
of $\gPin(2k)$, but this is inequivalent to the previous one because
the module eigenvalue of $\alpha$ on $|\psi \rangle$ is opposite. (The
latter may also be deduced as in~\cite{ref:Nee20b} from the
anticommutation of $\sigma$ with the module action of certain
reflection in $\gO(d)$.) It is convenient to assign highest weights
and Young diagrams as follows to irreducible $\gPin(2k)$
representations whose restrictions to $\lo(2k)$ are irreducible with
highest weights $\mu$ and such that the highest weight vectors of
these $\lo(2k)$ representations defined by identifying the
transformations on the left in~\eqref{eq:2k} with those on the right
for $d = 1$ are module eigenvectors of $\alpha$ with eigenvalues
$\pm 1$. If the eigenvalue is 1, the diagram is identical to the
$\lo(2k)$ diagram. If the eigenvalue is $-1$, the depth $\tilde \mu_1$
of the first column is increased to $2k - \tilde \mu_1$. The highest
weight is the set of row lengths of the diagram, where the rows are
labelled from the top by $\tau = k, k - 1, \dots, - k$ with $\tau = 0$
omitted and the length 0 is assigned to empty rows.

When $d$ is even and $\lambda_1 = 0$, equation \eqref{eq:sigD} gives
\begin{equation}
  \sigma \phi_\lambda = (-)^{\mu_1} \phi_{\lambda'} , \quad
  \sigma \phi_{\lambda'} = (-)^{\mu_1} \phi_\lambda ,
\end{equation}
where $\mu_1$ is the length of the lowest row in the diagram
complementary to $\lambda$ in Theorem~\ref{th:oo}.
When $d$ is odd, one gets
\begin{equation}
  \sigma \phi_\lambda = (-)^{\mu_1-1/2} \phi_{\lambda'} , \quad
  \sigma \phi_{\lambda'} =  (-)^{\mu_1+1/2}  \phi_\lambda .
\end{equation}
When the factor $(-)^{\mu_1}$ or $ (-)^{\mu_1-1/2}$ is included in
$\phi_{\lambda'}$, these maps become
\begin{equation}
  \sigma \phi_\lambda = \phi_{\lambda'} , \quad
  \sigma \phi_{\lambda'} = \phi_\lambda
\end{equation}
and
\begin{equation}
  \sigma \phi_\lambda = \phi_{\lambda'} , \quad
  \sigma \phi_{\lambda'} = - \phi_\lambda ,
\end{equation}
respective. By arguments similar to those spelled out above, it hence
follows that the $\lo(2k)$ module $\Psi_\mu$ of Theorem~\ref{th:oo}
carries an \emph{irreducible} representation of $\gPin(2k)$ with
$\lo(2k)$ highest weight vectors
$|\pm \mu \rangle \in \bar\Psi_{\pm\mu}$ such that $\alpha$ has the
module action $\alpha |\mu \rangle = |-\mu \rangle$,
$\alpha |-\mu \rangle = \pm |\mu \rangle$ according to whether $\mu$
has integral or half-integral components. I assign to a $\gPin(2k)$
representation composed of irreducible $\lo(2k)$ representations with
highest weights $\pm \mu$, where $\mu_1 > 0$, and with this module
action of $\alpha$, the Young diagram of $\mu$ and the highest weight
obtained by appending $k$ elements 0 at the front of $\mu$. Thus I
arrive at the following implication of Theorem~\ref{th:oo}.

\begin{thm}[$\lo(d)$--$\gPin(2k)$ duality]\label{th:oP}
  The fermion Fock space $\Phi$ has the decomposition
\begin{equation}\label{eq:oP}
  \Phi = \bigoplus {\textup X}_\lambda \otimes \Psi_\mu ,
\end{equation}
where $\textup{X}_\lambda$ and $\Psi_\mu$ carry irreducible
representations of $\lo(d)$ and $\gPin(2k)$ with highest weights
$\lambda$ and $\mu$. The sum in~\eqref{eq:oP} runs over all pairs of
$(\lambda,\mu)$ such that the $\lambda$ Young diagram and a reflected
and rotated copy of the $\mu$ Young diagram fill a $d/2 \times k$
frame without overlap when a negative $\lambda_1$ cancels a part of
the $\mu$ Young diagram extruding the frame as in the following
example, where $(d,k) = (12,4)$, $\lambda = (-1,1,2,2,3,4)$ and
$\mu = (0,0,0,1,1,2,4,5)$.
\begin{equation}\label{eq:o-Pin-fig}
  \includegraphics{young32}
\end{equation}
\end{thm}

\begin{rem}
  It is seen from~\eqref{eq:rho-1} that $\rho(-1) = 1$ for even $d$.
  The representation $\rho$ then factors through $\gO(2k)$. In this
  case the $\gPin(2k)$ Young diagrams are the usual $\gO(2k)$ Young
  diagrams~\cite{ref:Wey39}. For even $d$, Theorem~\ref{th:oP} is in
  fact obtained quickly from the theorem of $\gO(d)$-$\lo(2k)$
  duality~\cite{ref:Nee21,ref:Nee20b} by writing $a_{p\tau}$ and
  $a_{p\tau}^\dagger$ as $a_{-p,-\tau}^\dagger$ and $a_{-p,-\tau}$ for
  $p < 0$ so that the Fock space representation of $\lo(2k)$ becomes
  number conserving and that of $\lo(d)$ number non-conserving.
\end{rem}

\section{\label{sec:dis}Discussion}

Several authors discuss dual actions of finite- and
infinite-dimensional Lie algebras and, more generally, Lie
superalgebras on fermion or boson Fock spaces or combinations. A part
of this work is reviewed in~\cite{ref:Nee21,ref:Nee20b}. Other
references include~\cite{ref:Has89}, where, in particular,
Theorem~\ref{th:oo} is found. A systematic study of dual number
conserving actions of classical groups and number non-conserving
actions of Lie superalgebras, which include Lie algebras as special
cases, was initiated by Howe~\cite{ref:How89} (preprint 1976) and
continued in work including
\cite{ref:How95,ref:Wan99,ref:Che04,ref:Che10,ref:Che12}. The
\textit{classical groups} are the general linear groups $\gGL(n)$, the
orthogonal groups $\gO(n)$, the symplectic groups $\gSp(n)$ and their
subgroups. Some special cases of Howe's general duality
theorem~\cite{ref:How89} are presented in
~\cite{ref:Nee21,ref:Nee20b}. Evidently, every such duality implies by
restriction a duality involving the Lie algebra of the pertinent
group.

A basic tool in Howe's analysis~\cite{ref:How89,ref:How95} is a set of
results in classical invariant theory which Weyl calls the
\textit{first main theorems} of the classical groups~\cite{ref:Wey39}.
Howe and successors~\cite{ref:Goo98,ref:Che12} call them first
fundamental theorems. According to its first main theorem, the algebra
of invariants of a classical group is generated by its quadratic
invariants. Wang~\cite{ref:Wan99} suggests dual actions of groups
$\gPin(n)$ and certain infinite-dimensional Lie algebras, invoking
similarity with arguments in~\cite{ref:How95} which employ the
classical first main theorems. Similarly, Cheng, Kwon and
Wang~\cite{ref:Che10} propose dual actions of groups $\gPin(n)$ and
either Lie superalgebras or certain infinite-dimensional Lie algebras,
invoking similarity with arguments by Cheng and
Zhang~\cite{ref:Che04}, who, in turn, refer to~\cite{ref:How89}.
However, Howe does not list the groups $\gPin(n)$ among those with
known first main theorems~\cite{ref:How94}.

For comparison my proof of Theorem~\ref{th:oP} is based on
Theorem~\ref{th:oo}, which can be obtained without recourse to a first
main theorem. Anyway the purely fermionic case of Theorem A.1
in~\cite{ref:Che10} is equivalent to the odd $d$ case of my
Theorem~\ref{th:oP}. The representations considered there are
obtained, indeed, from the present ones by writing $a_{p\tau}$ and
$a_{p\tau}^\dagger$ as $a_{-p,-\tau}^\dagger$ and $a_{-p,-\tau}$ for
$p \le 0$. This renders neither representation of $\lo(d)$ nor
$\lo(2k)$ number conserving in general. My result thus provides a
proof of the purely fermionic case of the theorem in~\cite{ref:Che10}
that does not rely on a first main theorem.

\section{\label{sec:sum}Summary}

The result of my discussion is Theorem~\ref{th:oP}, which establishes
an orthogonal-Pin duality of Fock space representations. The proof is
based on Theorem~\ref{th:oo}, which was obtained in~\cite{ref:Nee19}
by a calculation of characters analogous to Helmers's proof of a
symplectic-symplectic duality~\cite{ref:Hel61}. Theorem~\ref{th:oP} is
closely analogous to the theorem of $\gO(d)$-$\lo(2k)$ duality
discussed in~\cite{ref:Nee21,ref:Nee20b} by establishing a 1--1
correspondence between representations of a Lie algebra and a group.
Both differ in this respect from Theorem~\ref{th:oo}, which
establishes a 2--1 or 1--2 correspondence between representations of
Lie algebras. Together, these three theorems present of remarkably
symmetric pattern. In~\cite{ref:Nee20b}, the theorem of
$\gO(d)$-$\lo(2k)$ duality is derived from Theorem~\ref{th:oo} like
Theorem~\ref{th:oP} was done above. Both theorems thus follow from
Theorem~\ref{th:oo}, and inversely, both of them obviously imply
Theorem~\ref{th:oo}. This renders the three theorems equivalent and
ultimately based on a calculation of characters. In this respect, the
proof of the theorem of $\gO(d)$-$\lo(2k)$ duality
in~\cite{ref:Nee20b} differs from a derivation as a special case of a
general duality theorem due to Howe whose proof is based on a set of
results in classical invariant theory which Weyl calls the first main
theorems on invariants of classical groups. (Howe and successors call
them first fundamental theorems.) Theorem~\ref{th:oP} could not be
obtained in this way due to a lack of first main theorems for Pin
groups. My result provides a proof that does not rely on such a
theorem of the purely fermionic case of a theorem on dual Fock space
representations of a Lie superalgebra and a Pin group formulated by
Cheng, Kwon and Wang.

\bibliography{pin2}

\end{document}